\documentclass[preprint]{revtex4-1} 

\usepackage{amsfonts} 
\usepackage{graphicx}
\usepackage{amsmath, amssymb}
\usepackage{slashed}
\usepackage{braket}
\usepackage{physics}
\usepackage[bottom]{footmisc}
\graphicspath{  {images/} }
\begin{document}


\title{Coherent States and Generalized Hermite Polynomials for Fractional statistics - interpolating from fermions to bosons}
\author{Satish Ramakrishna}
\email{ramakrishna@physics.rutgers.edu} 


\date{\today}

\begin{abstract}

This article develops the geometric structure that results from the $\theta$-commutator  $\alpha \beta - e^{i \theta} \beta \alpha = 1 $  that provides a continuous interpolation between the Clifford and Heisenberg algebras. We first demonstrate the most general geometrical picture, applicable to all values of $N$. After listing the properties of this Hilbert space, we study the calculus of generalized coherent states that result  when $\xi^N=0$, for $N \ge 2$, including a calculation of the free-energy for particles of intermediate statistics. Lastly, we solve the generalized harmonic oscillator problem and derive generalized versions of the Hermite polynomials for general $N$.

Some remarks are made to connect this study to the case of anyons. This study represents the first steps towards developing an anyonic field theory.

\end{abstract}

\maketitle 

\section{Introduction \& Motivation}
In an earlier paper\cite{Satish1}, we analyzed the Hilbert space derived from the ``commutator''
\begin{eqnarray}
\alpha \beta - e^{i \theta} \beta \alpha = 1 
\end{eqnarray}
in the case where $\theta = \frac{2 \pi M}{N}$ where $M, N$ are co-prime non-zero natural numbers and where we also define $z=e^{i \theta}$. This algebra was inspired by the properties of anyons and was intended to interpolate between fermionic and bosonic statistics.
In particular, we had considered the general case where the vacuum state had a non-zero eigenvalue. When the vacuum state has a zero eigenvalue, however, we are naturally led to the study of variables $\xi$, such that $\xi^N=0$. Such variables, which may be referred to as generalized Grassmann variables have been the subject of much study in the past \cite{Biedenharn, MacFarlane, Chaichian}, of which the most complete and relevant is \cite{Chaichian}. The difference is that since our focus is on the full range of fractional statistics between fermions and bosons in 2+1 dimensions, we study a full calculus starting with integration rules to a physically reasonable construction of the path integral for the free energy.

In addition to the generalization of the Grassmann variables mentioned above, there is also a rather simple geometrical picture that emerges from the arithmetic for the operators in the algebra \cite{Hallnas}. Akin to the fuzzy solid representations used for the angular momentum algebra, we prove that the relevant geometrical picture here is a pancake that goes from a sphere (for $N=2$) to a plane (for $N \rightarrow \infty$).

\section{Summary of Properties}
We are going to, in this paper, analyze several properties of the Hilbert space, in the special and physically interesting case where the vacuum state has zero eigenvalue for the operators $\beta \alpha$, as well as $\alpha^{\dagger} \alpha$. Hence, in the notation of \cite{Satish1}, we set $\lambda_0=0$.

For general integer $N$, we deduce the following properties.
\begin{enumerate}
\item States are labeled by their eigenvalues under $\beta \alpha$, i.e., 
\begin{eqnarray}
\beta \alpha \ket{\lambda_m} = \lambda_m \ket{\lambda_m} \nonumber \\
Eigenvalues: \: \lambda_0 \equiv \lambda_{N}=0, \lambda_1=1, \lambda_2=1+z,\lambda_3=1+z+z^2, \:  ... \: , \nonumber \\
 \lambda_{N-1}=1+z+z^2+...+z^{N-2}, 
\: and \: \: \lambda_{N}=1+z+z^2+...+z^{N-1}=0 \nonumber \\ 
Eigenstates: \: \ket{\lambda_0} \equiv \ket{0}, \ket{\lambda_1}, \ket{\lambda_2}, ...\ket{\lambda_{N-1}}, \: \: \: \:  \: \: \: \: \nonumber \\
\ket{\lambda_{N}}\equiv \ket{\lambda_0} \equiv \ket{0}  \: \: \: \:  \: \: \: \: \nonumber \\
\braket{\lambda_m}{\lambda_n}=\delta_{nm} \: \: \: \:  \: \: \: \:  \: \: \: \: 
\end{eqnarray}
The eigenvectors can be constructed to be orthogonal, since these are also the eigenvectors of the usual number operator $\alpha^{\dagger} \alpha$.

We propose to call these states ``overons'', since they represent one of the two ways to flip anyons (``over'' and ``under''). The complex conjugate states would be then called ``underons''. In Appendix 1, we study possible dynamical system analogs that might result from such excitations.
\item The actions of the operators are
\begin{eqnarray}
\alpha \ket{0}\equiv \alpha \ket{\lambda_{N}} \equiv \alpha \ket{\lambda_0}= 0 \nonumber \\
\alpha \ket{\lambda_m} = \sqrt{\lambda_m} \ket{\lambda_{m-1}} \: \: \: \: \: \: m>0\nonumber \\
\beta \ket{\lambda_{m-1}} = \sqrt{\lambda_m} \ket{\lambda_m}
\end{eqnarray}
\item A consistent identification is $\beta = \alpha^T, \: \alpha= \beta^T$, i.e., the commutator is $\alpha \alpha^T - z \alpha^T \alpha = 1$..
\item When we take the complex conjugate of the basic commutator, we get, by an entirely similar procedure to the above, that the eigenstates of $a^{\dagger} a^* $ are $\ket{\lambda_m^*}$. The chain of reasoning is
\begin{eqnarray}
\alpha \alpha^T - z \alpha^T \alpha = 1 \nonumber \\
\alpha^* \alpha^{\dagger} - z^{-1} \alpha^{\dagger} \alpha^* = 1 \nonumber \\
(\alpha^T \alpha) \ket{\lambda_m} = \lambda_m \ket{\lambda_m} \nonumber \\
\alpha^{\dagger} \alpha^* \ket{\lambda^*_m} = \lambda_m^* \ket{\lambda_m^*}
\end{eqnarray}
Taking the complex conjugate of the third equation above, we get $\left( \ket{\lambda_m} \right)^* = \ket{\lambda_m^*}$.
This leads to the equations
\begin{eqnarray}
\alpha \ket{\lambda_m} = \sqrt{\lambda_m} \ket{\lambda_{m-1}} \nonumber \\
\alpha^* \ket{\lambda_m^*} = \sqrt{\lambda_m^*} \ket{\lambda_{m-1}^*}
\end{eqnarray}
Continuing, we can now write $\ket{\lambda_m^*}$ as a linear combination of the $\ket{\lambda_m}$. In fact, it is easy to see, from the geometry of the eigenvectors on the complex plane in Fig. 3, that
\begin{eqnarray}
\ket{\lambda_m^*} = z^{-(m-1)} \ket{\lambda_m}
\end{eqnarray}
Using this,
\begin{eqnarray}
\alpha \ket{\lambda_m^*} = z^{- (m-1)} \sqrt{\lambda_m} \ket{\lambda_m} = \sqrt{\lambda_m} \ket{\lambda_m^*} \: \:  \: \: \: \: \: \: \:  \: \: \: \: \: \nonumber \\
\alpha^{\dagger} \ket{\lambda_m} = (\alpha^T)^* \ket{\lambda_m} = \left( \alpha^T \ket{\lambda_m^*}\right)^* = \left(  z^{-(m-1)} \alpha^T \ket{\lambda_m}\right)^* = z^{m-1} \sqrt{\lambda_{m+1}^*} \ket{\lambda_{m+1}^*} \nonumber \\
= z^{-1} \sqrt{\lambda_{m+1}^*} \ket{\lambda_{m+1}}  \: \: \: \: \:  \: \: \: \: \:  \: \: \: \: \: \: \:  \: \: \: \: \: \: \:  \: \: \: \: \: 
\end{eqnarray}
Using this, and defining $\bra{\lambda_m}$ as the usual hermitian conjugate transpose of $\ket{\lambda_m}$,
\begin{eqnarray}
\alpha \alpha^{\dagger} \ket{\lambda_m} = |\lambda_{m+1}| \ket{\lambda_m} \: \:  \: \: \: \: \:  \nonumber \\
\alpha^{\dagger} \alpha \ket{\lambda_m} = |\lambda_m| \ket{\lambda_m} \nonumber \\
\rightarrow \: \: \bra{\lambda_m} \alpha \alpha^{\dagger} - \alpha^{\dagger} \alpha \ket{\lambda_m} = |\lambda_{m+1}| - |\lambda_m|
\end{eqnarray}
The traditional ``number'' operator $\alpha^{\dagger} \alpha$ is diagonal in the same basis that $\alpha^T \alpha$ is. Since $\alpha^{\dagger} \alpha$ is a hermitian operator, it is consistent that the $\ket{\lambda_m}$ is an orthonormal basis \cite{Banks}.
\item The eigenvalue spectrum of $\beta \alpha \equiv \alpha^T \alpha$ (magnitude as well as complex vectors) is as below and in reference \cite{Satish1} and is plotted in Fig. 1. Note that we have $\theta = \frac{2 \pi M}{N}$ and we have used $M=1$ for the graphs in Fig. 1.
\begin{eqnarray}
\lambda_0=0 \: \:  \: \: \: \: \:  \: \:  \: \: \: \: \: \: \:  \: \: \: \: \: \: \:  \: \: \: \: \:    \nonumber \\
\lambda_m = z^0+z^1 +...+z^{m-1}=\frac{1 - e^{i m \theta}}{1 - e^{i \theta}} = e^{i \frac{(m-1) \theta}{2}}  \frac{\sin{\frac{m\theta}{2}}}{\sin{\frac{\theta}{2}}}
\end{eqnarray}
\begin{figure}[h!]
\caption{Eigenvalue Spectrum}
\centering
\includegraphics[scale=.36]{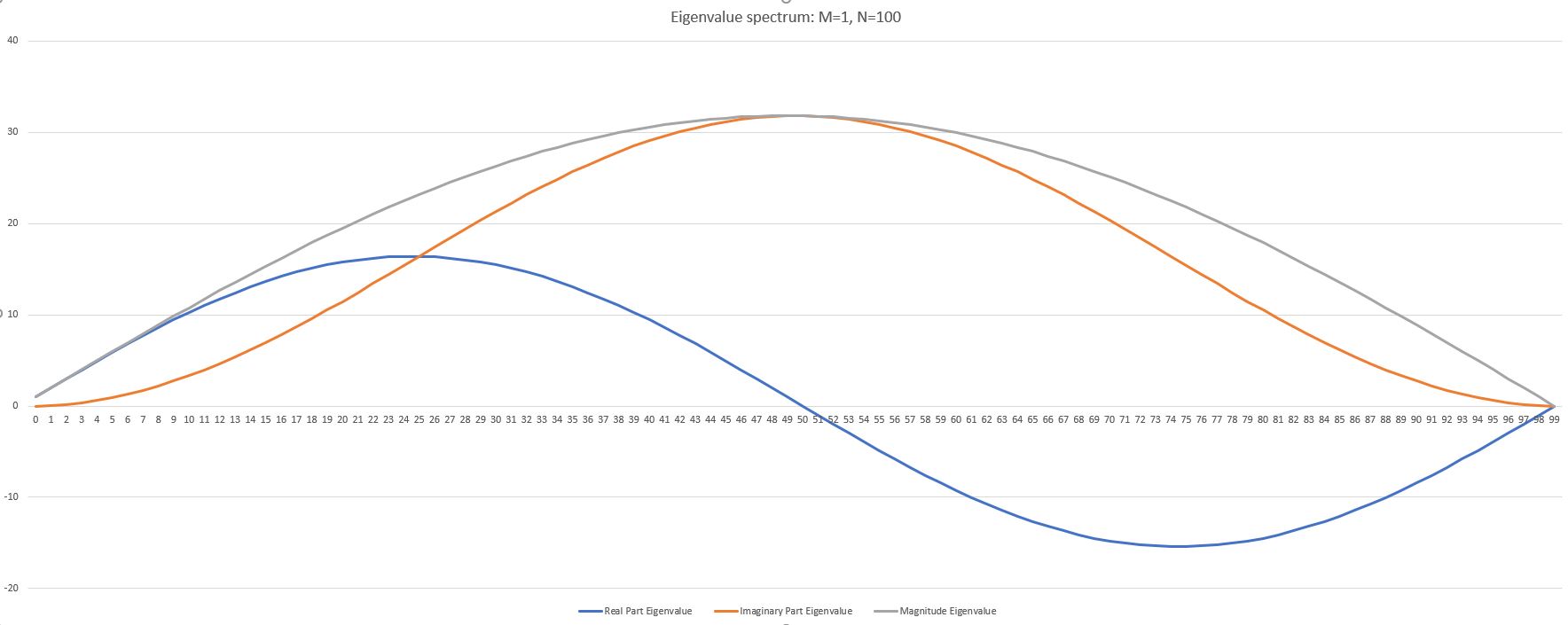}
\includegraphics[scale=0.45]{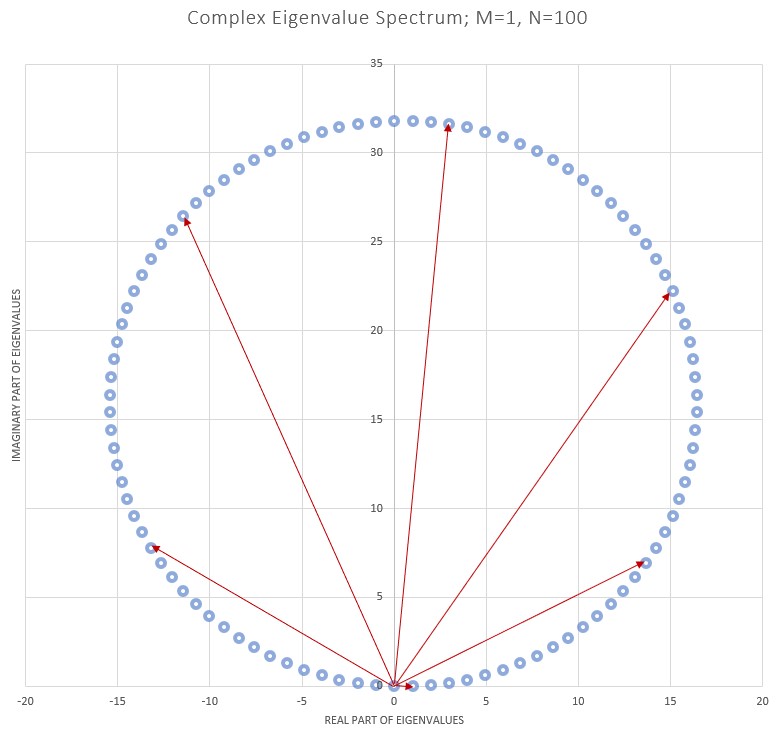}
\end{figure}
\item The matrices $\alpha$ and $\beta = \alpha^T$ are displayed explicitly below, for $\theta = \frac{2 \pi M}{N}$. They are $N \times N$ matrices, since the eigenstates are $N$-dimensional vectors. The eigenvectors are as below
\begin{eqnarray}
\ket{\lambda_0} = \left(\begin{array}{c} {\bf 1} \\
					     0 \\
					     0  \\
					     0 \\
					     . \\
					     . \\
					     . \\
					     0  \\
					     0  \end{array}\right) \nonumber \: \: \: , \: \:
					     \ket{\lambda_1} = \left(\begin{array}{c} 0 \\
					     {\bf 1} \\
					     0  \\
					     0 \\
					     . \\
					     . \\
					     . \\
					     0  \\
					     0  \end{array}\right)  \: \: \: , \: \: 
					     \ket{\lambda_2} = \left(\begin{array}{c} 0 \\
					     0 \\
					     {\bf 1}  \\
					     0 \\
					     . \\
					     . \\
					     . \\
					     0  \\
					     0  \end{array}\right)   \: \: \: ...  \: \: 
					     \ket{\lambda_{N-1}} = \left(\begin{array}{c} 0 \\
					     0 \\
					     0  \\
					     0 \\
					     . \\
					     . \\
					     . \\
					     0  \\
					     {\bf 1} \end{array}\right)  \: \: \: \: \: \: \: \: \: \:  \: \: \: \: \: \: \: \: \: \:  \: \: \: \: \: \: \: \: \: \: 
\end{eqnarray}
while the operators are
\begin{eqnarray}
\alpha=
\left(\begin{array}{cccccccc} {\bf 0} & \: \: \: 1 &\: \: \:  0 &\: \: \:  0 &\: \: \:  0 & ... & \: \: \:  0 &0\\
					     0 & \: \: \: {\bf 0} & \: \: \: \sqrt{1+z} & \: \: \:  0 & \: \: \:  0 & ... & \: \: \:  0 &0 \\
					     0 & \: \: \: 0 & \: \: \: {\bf 0} &\: \: \: \sqrt{1+z+z^2} & \: \: \:  0 &... & \: \: \:  0 &0 \\
					     0 & \: \: \: 0 & \: \: \: 0 &\: \: \: {\bf 0} & \: \: \:  \sqrt{1+z+z^2+z^3} &... & \: \: \:  0 &0 \\
					     . \\
					     . \\
					     . \\
					     0 & \: \: \: 0 & \: \: \: 0 &\: \: \:  0 & \: \: \:  0 & ... &\: \: \:  {\bf 0}  & \sqrt{1+z+z^2+...+z^{N-1}}  \\
					     0 & \: \: \: 0 & \: \: \: 0 &\: \: \:  0 & \: \: \:  0 & ... &\: \: \:  0 & {\bf 0} \end{array}\right)
					     \nonumber \\
					     \beta = 
					     \alpha^T=
\left(\begin{array}{cccccccc} {\bf 0} & \: \: \: 0 &\: \: \:  0 &\: \: \:  0 & \: \: \:  0 & ... & \: \: \:  0 & 0\\
					     1 & \: \: \: {\bf 0} & \: \: \: 0 & \: \: \:  0 & \: \: \:  0 & ... &\: \: \:  0 & 0 \\
					     0 & \: \: \: \sqrt{1+z} & \: \: \: {\bf 0} &\: \: \: 0 & \: \: \:  0 & ... &\: \: \:  0 & 0 \\
					     0 & \: \: \: 0 & \: \: \: \sqrt{1+z+z^2} & \: \: \: {\bf 0}& \: \: \:  0 & ... &\: \: \:  0 & 0\\
					      0 & \: \: \: 0 & \: \: \: 0 &\: \: \: \sqrt{1+z+z^2+z^3} & \: \: \: {\bf 0}&  ... &\: \: \:  0 & 0\\
					     . \\
					     . \\
					     . \\
					     0 & \: \: \: 0 & \: \: \: 0 &\: \: \:  0 & \: \: \:  0 & ... & \sqrt{1+z+z^2+...+z^{N-1}} &  {\bf 0} \end{array}\right) \nonumber
\end{eqnarray}
while the commutator is
\begin{eqnarray}
					     \alpha \alpha^T - \alpha^T \alpha = \alpha \beta - \beta \alpha = \left(\begin{array}{cccccccc} {\bf 1} & \: \: \: 0 &\: \: \:  0 &\: \: \:  0 & \: \: \:  0 & ... & 0\\
					     0 & \: \: \: {\bf z} & \: \: \: 0 & \: \: \:  0 & \: \: \:  0 & ... &  \: \: \:  0 & 0 \\
					     0 & \: \: \: 0 & \: \: \: {\bf z^2} &\: \: \: 0 & \: \: \:  0 & ... &  \: \: \:  0 & 0 \\
					     0 & \: \: \: 0 & \: \: \: 0 & \: \: \: {\bf z^3}& \: \: \:  0 & ... &  \: \: \:  0 & 0\\
					      0 & \: \: \: 0 & \: \: \: 0 &\: \: \: 0 & \: \: \: {\bf z^4}& ... &  \: \: \:  0 & 0\\
					     . \\
					     . \\
					     . \\
					     0 & \: \: \: 0 & \: \: \: 0 &\: \: \:  0 & \: \: \:  0 & ... &  {\bf z^{N-2}} &  \: \: \:  0 \\
					     0 & \: \: \: 0 & \: \: \: 0 &\: \: \:  0 & \: \: \:  0 & ... &  \: \: \:  0 &  {\bf z^{N-1}} \end{array}\right)  \: \: \: \: \: \: \: \: \: \: \: \:  \nonumber \\
					     = diag \left[1,z,z^2,z^3, ... , z^{N-1} \right]  \: \: \: \: \: \: \: \: \: \: \: \: \: \: \: \: \: \: \: \: \: \: \: \: \: \: \: \: \: \: \: \: \: \: \: \: \: \: \: \: \: \: \: \: \: \: \: \: \: \: \: \: \: \: \: \: \: \: \: \: \: \: \: \: \: \: \: \: \: \: \: \:
\end{eqnarray}
and
\begin{eqnarray}
\alpha \alpha^{\dagger} - \alpha^{\dagger} \alpha = diag \left[1-0,|\lambda_2|-|\lambda_1|,|\lambda_3|-|\lambda_2|, ... , |\lambda_N|-|\lambda_{N-1}| \right]  \nonumber \\
=diag[1,\frac{\sin \frac{2 \theta}{2} - \sin \frac{\theta}{2}}{\sin \frac{\theta}{2}},\frac{\sin \frac{3 \theta}{2} - \sin \frac{2 \theta}{2}}{\sin \frac{\theta}{2}},\frac{\sin \frac{4 \theta}{2} - \sin \frac{3\theta}{2}}{\sin \frac{\theta}{2}}, ... , \frac{- \sin \frac{(N-1)\theta}{2}}{\sin \frac{\theta}{2}}]  \: \: \: \: \: \: \: \: \: \: \: \: \: \: \: \: \: \:
\end{eqnarray}
and
\begin{eqnarray}
\alpha \alpha^{\dagger} - {\mathcal C} \alpha^{\dagger} \alpha = 1 \nonumber \\
{\mathcal C} = diag[ \frac{|\lambda_1| - 1}{|\lambda_0|},\frac{|\lambda_2| - 1}{|\lambda_1|},\frac{|\lambda_3| - 1}{|\lambda_2|}, ..., \frac{|\lambda_{N}| - 1}{|\lambda_{N-1}|}]
\end{eqnarray}

The top-left component $\frac{|\lambda_1| - 1}{|\lambda_0|}$ of $\mathcal C$ is not determined by the above commutator since $\alpha^{\dagger} \alpha=0$ for the state $\ket{\lambda_0}$. However, we can determine it by taking the limit of the expressions for $\lambda_0 \rightarrow 0$ as in \cite{Satish1}; it becomes $\cos \theta$, which is $-1$ for the fermion limit and $+1$ for the bosonic limit. Hence $\mathcal C$ is $- \cal I$ (the identity matrix) for fermions and $+\cal I$ for bosons.

\item When we compute scattering amplitude matrix elements for different particles, we will have to re-order the annihilation and creation operators, then will be left with a product of terms like $\alpha^m \beta^m$, i.e.,
\begin{eqnarray}
\alpha \beta = (1+z \: \beta \alpha) \nonumber \\
\alpha^2 \beta^2 = (1+z) + (z+2 z + z^3) \beta \alpha + z^4 \beta^2 \alpha^2 \: \nonumber \\
... \nonumber 
\end{eqnarray}
When we compute expectation values in the vacuum state (i.e., $\bra{0}\alpha^m \beta^m \ket{0}$), only the constant terms will be left and they are, for the first few powers
\begin{eqnarray}
m=1 \: \: \: \rightarrow \: \: \: 1 \nonumber \\
m=2 \: \: \: \rightarrow \: \: \: 1+z \nonumber \\
m=3 \: \: \: \rightarrow \: \: \: 1+2z+2z^2+z^3 \nonumber \\
m=4 \: \: \: \rightarrow \: \: \: 1+3z+5z^2+6z^3+5z^4+3z^5+z^6 \nonumber \\
m=5 \: \: \: \rightarrow \: \: \: 1+4z+9z^2+15z^3+20z^4+22z^5+20z^6+15z^7+9z^8+4z^9+z^{10} \nonumber \\
m=6 \: \: \: \rightarrow (1,5,14,29,49,71,90,101,101,90,71,49,29,14,5,1) \: \: \: \: \: \: \: \: \: \: \: \: \: \: \: 
\end{eqnarray}
where we have represented the polynomial by just its coefficients (the Mahonian numbers \cite{Mahon}) in the last case. As can be checked quickly, these are the polynomials $\lambda_1, \lambda_2 \times \lambda_1, \lambda_3 \times \lambda_2 \times \lambda_1$ etc.
\end{enumerate}

\section{Geometrical Interpretation}

The most general geometrical construction is
\begin{eqnarray}
X=\frac{\alpha +\alpha^{\dagger}}{2} \; , \; Y = \frac{\alpha -\alpha^{\dagger}}{2i} \: , \: 
Z = \frac{1}{2} [ \alpha, \alpha^{\dagger} ]  \nonumber \\
X^2+Y^2 = \frac{1}{2} \left( \alpha \alpha^{\dagger} + \alpha^{\dagger} \alpha \right) \nonumber \\
2 Z = \left( \alpha \alpha^{\dagger} - \alpha^{\dagger} \alpha \right) \nonumber \\
{\cal L}_{m,n} \equiv \left( \alpha \alpha^{\dagger} + \alpha^{\dagger} \alpha \right)_{m,n} = \left( |\lambda_{m+1}| + |\lambda_m|\right) \delta_{m,n} \nonumber \\
{\cal M }_{m,n} \equiv \left( \alpha \alpha^{\dagger} - \alpha^{\dagger} \alpha \right)_{m,n} = \left( |\lambda_{m+1}| - |\lambda_m|\right) \delta_{m,n}
\end{eqnarray}
and observe that 
\begin{eqnarray}
\bigg( (|\lambda_{m+1}| + |\lambda_m|) \sin \frac{\theta}{4} \bigg)^2+\bigg( (|\lambda_{m+1}| - |\lambda_m|) \cos \frac{\theta}{4} \bigg)^2 =1
\end{eqnarray}
we obtain the equation
\begin{eqnarray}
(X^2+Y^2)^2 \sin^2 \frac{\theta}{4} + Z^2 \cos^2 \frac{\theta}{4} = \frac{1}{4}
\end{eqnarray}
This equation can be re-phrased as an invariant of the group that underlies the algebra. The algebra can be written most simply with the creation/annihilation operators as
\begin{eqnarray}
\left[\alpha, \alpha^{\dagger} \right] = 2 Z \nonumber \\
\left[\alpha, Z \right] = S_D \: \alpha \nonumber \\
\left[\alpha^{\dagger}, Z \right] = - \alpha^{\dagger} \: S_D \nonumber \\
\tan^2 \frac{\theta}{4} = \frac{1 - {\cal M}^2}{{\cal L}^2-1}
\end{eqnarray}
In Equation (17), the two limits $\theta \rightarrow 0$ and $\theta \rightarrow \pi$ are consistent on both sides of the equation.
Additionally, ${\cal L}$ and ${\cal M}$ are the matrices as defined in Equation (23) and $S_D$ is the real, diagonal matrix
\begin{eqnarray}
S_D = diag \left[\frac{|\lambda_0|}{2}+\frac{|\lambda_2|}{2}-|\lambda_1|,\frac{|\lambda_1|}{2}+\frac{|\lambda_3|}{2}-|\lambda_2|, ... , \frac{|\lambda_{N-2}|}{2}+\frac{|\lambda_N|}{2}-|\lambda_{N-1}| \right]  \
\nonumber \\
\equiv D_m^2 (|\lambda|)\: \: \: \: \: \: \: \: \: \: \: \:  \: \:  \: \:  \: \: 
\end{eqnarray}
which is the finite-difference Laplacian of a diagonal matrix with absolute values of the eigenvalues along the diagonal. In this notation, from Equation (17), $Z = D_m(|\lambda|)$, so that the commutator in that equation can be written in the interesting form
\begin{eqnarray}
\left[\alpha, D_m(|\lambda|) \right] = D_m^2(|\lambda|) \alpha
\end{eqnarray}

Incidentally, for $N=3$, this is consistent with the previous paper's\cite{Satish1}  Equation (25) at the points resolved within the fuzzy ellipsoid ($Z=J_z=0,\pm \frac{1}{2}$). In this case,
\begin{eqnarray}
J_z = \frac{1}{2} \left(\begin{array}{ccc} {\bf 1} & 0 & 0\\
0 & {\bf 0} & 0 \\
0 & 0 & {\bf -1}  \end{array}\right) \: , \: J_x = \frac{\alpha +\alpha^{\dagger}}{2} = \frac{1}{2} \left(\begin{array}{ccc} 0 & {\bf 1} & 0\\
{\bf 1} & 0 & {\bf \sqrt{1+e^{i \frac{2 M \pi}{3}}}} \\
0 & {\bf \sqrt{1+e^{-i \frac{2 \pi M}{3}}}}  & 0 \end{array}\right) \: , \nonumber \\
\: J_y = \frac{\alpha -\alpha^{\dagger}}{2i} = \frac{i}{2} \left(\begin{array}{ccc} 0 & {\bf -1} & 0\\
{\bf 1} & 0 & -\bf{\sqrt{1+e^{i \frac{2 M \pi}{3}}}} \\ 
0 & \bf{\sqrt{1+e^{-i \frac{2 \pi M}{3}}}} &  0 \end{array}\right) \nonumber \\
J_z^2 = \frac{1}{4} \left(\begin{array}{ccc} {\bf 1} & 0 & 0\\
0 & {\bf 0} & 0 \\
0 & 0 & {\bf 1}  \end{array}\right) \: , \: J_x^2+J_y^2  = \frac{1}{4} \left(\begin{array}{ccc} {\bf 2}  & 0 & 0\\
0 & {\bf 4} & 0 \\
0 & 0 & {\bf 2} \end{array}\right) \: \: \: \: \: \: \: \: \: \: \: \: \: 
\end{eqnarray}
satisfies both the Equations (25) and following in the previous paper \cite{Satish1}, i.e., 
\begin{eqnarray}
J_x^2+J_y^2+2 J_z^2 = 1 \: \: \: \:  \& \: \: \: \: (J_x^2+J_y^2)^2+3 J_z^2 = 1
\end{eqnarray}

We have already noted the equivalence for $N=2$.

The surface described by the above equation is pancake-shaped aligned along the z-axes, as in Fig. 2.

\begin{figure}[h!]
\caption{Geometrical Interpretation of the eigenvalue surface}
\centering
\includegraphics[scale=.5]{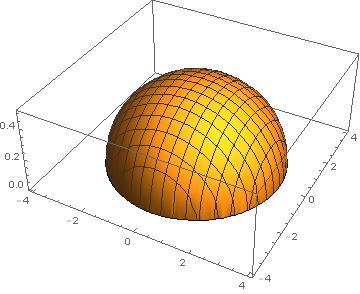}
\end{figure}

For large $N$, we can write the solutions to the above equation (in terms of the polar radius coordinate $\rho$) as
\begin{eqnarray}
\rho^2 = X^2+Y^2 = (m+\frac{1}{2})- \frac{mq^2}{12} (1+3m+2m^2)  \nonumber \\
z = \frac{1}{2}-\frac{mq^2}{4}(1+m) \: \: \: \: \: \: \: \: \: \: \: \: \: \: \: 
\end{eqnarray}
which yields concentric circular strips on the plane $z=\frac{1}{2}$, where each state has radius $\propto \sqrt{m+\frac{1}{2}}$. This is the usual picture for Landau levels; here this is a geometrical representation of the usual bosonic levels.

To show how a spherical object for $N=2$ transforms into the flat plane in the $N \rightarrow \infty$ limit, we can compute the surface area of the pancake. The area integral is computed (define $a=\sin \frac{\theta}{4}$) as
\begin{eqnarray}
{\cal A}(a) = 2 (2 \pi) \int_{\rho = 0}^{\rho = \frac{1}{\sqrt{2 a}}} \: d \rho \: \rho \sqrt{1 + (\frac{dz}{d \rho})^2} \nonumber \\
= \frac{2 \pi}{a} \int_{l=0}^{l=\frac{1}{2}} \: dl \sqrt{1 + \frac{4 a l^3}{(1-a^2)(\frac{1}{4} - l^2)}}
\end{eqnarray}
Clearly, this diverges as $a \rightarrow 0$, which corresponds to $N \rightarrow \infty$, since $\theta = \frac{2 \pi}{N}$.

We plot the area as a function of $a$ in Fig. 3. Indeed, the pancake like closed surface turns into the infinite plane as $N \rightarrow \infty$.

\begin{figure}[h!]
\caption{Area of the geometrical surface}
\centering
\includegraphics[scale=.5]{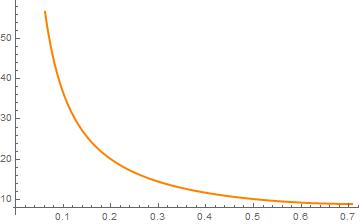}
\end{figure}

\section{Coherent States for general $N$ - Generalized Grassmann variables}

For a general $N$-type algebra, we can write down an eigenstate for $\alpha$ as
\begin{eqnarray}
 \alpha \ket{\xi} = \xi \ket{\xi} \: \: \: \: \: \: \: \: \: \: \: \: \: \: \: \: \: \: \: \: \: \: \: \: \: \: \: \: \: \: \: \: \: \: \: \: \: \: \: \: \: \: \: \: \: \: \: \: \: \: \: \: \: \: \: \: \: \: \: \: \: \: \: \: \: \: \: \: \: \: \: \: \: \: \: \: \: \: \: \: \: \: \: \: \: \: \: \: \: \: \: \: \: \: \: \: \: \nonumber \\
\ket{\xi} = \ket{\lambda_{0}}+ \frac{\xi}{\sqrt{\lambda_1}} \ket{\lambda_1} +\frac{\xi^2}{\sqrt{\lambda_2 \lambda_1}} \ket{\lambda_2} + \frac{\xi^3}{\sqrt{\lambda_3\lambda_2\lambda_1}} \ket{\lambda_3} + \: \: \: \: \: \: \: \: \: \: \: \: \: \: \: \: \:  \nonumber \\
 ... + \frac{\xi^{N-2}}{\sqrt{\lambda_{N-2} \lambda_{N-3}...\lambda_1}} \ket{\lambda_{N-2}} + \frac{\xi^{N-1}}{\sqrt{\lambda_{N-1}\lambda_{N-2}...\lambda_1}} \ket{\lambda_{N-1}} \: \: \: \: \: \: \: \: \: 
 \end{eqnarray}
 The above expansion is identical to the usual formula for boson coherent states in the limit $N \rightarrow \infty$, as well as for fermion coherent states at $N=2$ (the usual Grassmann variables). We also define the ``bra'' vector as the adjoint vector, where $\xi^{\dagger}$ is the adjoint of $\xi$ and is independent of $\xi$, i.e.,
 \begin{eqnarray}
 \bra{\xi} = \bra{\lambda_{0}}+ \frac{\xi^{\dagger}}{\sqrt{\lambda_1^*}} \bra{\lambda_1} +\frac{(\xi^{\dagger})^2}{\sqrt{\lambda_2^* \lambda_1^*}} \bra{\lambda_2} + \frac{(\xi^{\dagger})^3}{\sqrt{\lambda_3^* \lambda_2^* \lambda_1^*}} \bra{\lambda_3} + \: \: \: \: \: \: \: \: \: \: \: \: \: \: \: \nonumber \\
 ... + \frac{ (\xi^{\dagger})^{N-2}}{\sqrt{\lambda_{N-2}^* \lambda_{N-3}^*...\lambda_1^*}} \bra{\lambda_{N-2}} + \frac{(\xi^{\dagger})^{N-1}}{\sqrt{\lambda_{N-1}^* \lambda_{N-2}^*...\lambda_1^*}} \bra{\lambda_{N-1}} \nonumber \\
 \bra{\xi} \alpha^{\dagger} = \bra{\xi} \xi^{\dagger} \: \: \: \: \: \: \: \: \: \: \: \: \: \: \: \: \: \: \: \: \: \: \: \: \: \: \: \: \: \: \: \: \: \: \: \: \: \: \: \: \: \: \: \: \: \: \: \: \: \: \: \: \: \: \: \: \: \: \: \: \: \: \: \: \: \: \: \: \: \: \: \: \: \: \: \: \: \: \: \: \: \: \: \: \: \: \: \: \: \: \: \: \: \: \: \: \:
\end{eqnarray}
That they are eigenvectors may be checked by applying $\alpha$ to the state and using Equation (3), we postulate that $\xi$ and $\alpha$ commute. Additionally, we posit that $\xi^N=0$ and the transposition relations (the matrix ${\cal C}_0$ is as defined in Equation (12))
\begin{equation}
(\xi^{\dagger})^{\dagger}=\xi \: \: , \: \: \xi  \xi^{\dagger}= {\cal C}_0 \: \xi^{\dagger} \xi \: \: , \: \: \xi^{\dagger}\xi  = {\cal C}_0^{-1} \:  \xi\: \xi^{\dagger}
\end{equation}
What sort of object is $\xi$? The two relations above make sense only if $\xi$ were itself an $N \times N$ matrix, for consistency, we assume is a direct product of a ``generalized Grassmann matrix'' and a unit $N \times N$ matrix in the states' eigenbasis, hence, commutes with all other complex number matrices in the same eigenbasis. Hence $\xi^T$ and $\xi^{\dagger}$ are all reasonable objects to define. The above statements and equations are also consistent with the statement that a term like $\xi^{\dagger} \xi$ is ``real''. This is true, since $\cal C$ is a real matrix. In addition, since $\xi \sim \alpha$, it is consistent to require the equations below (in line with Equation (4)),
\begin{eqnarray}
\xi \xi^T = z \xi^T \xi \nonumber \\
\xi \xi^{\dagger} = {\cal C}_0 \xi^{\dagger} \xi \nonumber \\
\end{eqnarray}
The description of this algebra is similar to the treatment in reference \cite{Chaichian}, however, the difference here is that these variables are directly coherent state variables, as in the usual definition \cite{Murayama1}.

Some auxiliary results are written below. Again, note that all these products are scalars times the unit matrix.
\begin{eqnarray}
\braket{\xi}{\xi} = 1 + \frac{\xi^{\dagger} \xi }{|\lambda_1|} +  \frac{(\xi^{\dagger})^2 \xi^2  }{|\lambda_2 \lambda_1|} + \frac{(\xi^{\dagger})^3 \xi^3 }{|\lambda_3 \lambda_2 \lambda_1|} + ... +  \frac{(\xi^{\dagger})^{N-1} \xi^{N-1}  }{|\lambda_{N-1} \lambda_{N-2} ...\lambda_1|}
\end{eqnarray}
Note that in the $N \rightarrow \infty$ limit, when $z \rightarrow 1$, $\braket{\xi}{\xi}  = e^{ \xi^{\dagger} \xi}$, which is appropriate for bosonic coherent states. Taking the limit in the opposite direction, the behavior is appropriate for fermions, when $N=2$ the algebra automatically yields $\braket{\xi}{\xi}  = e^{ \xi^{\dagger} \xi} =  e^{- \xi  \xi^{\dagger}}$.

We can now write down, from the expansion in Equation (28), a series expansion for $\frac{1}{\braket{\xi}{\xi}}$, which also terminates at the term $(\xi^{\dagger})^{N-1} \xi^{N-1}$, since higher powers are $0$.

The following integrals are postulated, in order to match the boundary cases for bosonic variables as well as for fermionic $(N=2)$ Grassmanns. We set
\begin{eqnarray}
\int d \xi^{\dagger} d \xi \frac{1}{\braket{\xi}{\xi}}  = 1 \: \: \:  \: \: Normalization \: \: \nonumber \\
\int d \xi^{\dagger} d \xi \frac{1}{\braket{\xi}{\xi}}  \xi^{\dagger} \xi= {\cal C}_0 |\lambda_1|  \: \: \:  \: \: First \: Moment \: \: \nonumber \\
\int d \xi^{\dagger} d \xi \frac{1}{\braket{\xi}{\xi}}  (\xi^{\dagger})^2  \xi^2 = {\cal C}_0^{2} |\lambda_2 \lambda_1|  \: \: \:  \: \: Second \: Moment \: \: \nonumber \\
... \nonumber \\
\int d \xi^{\dagger} d \xi \frac{1}{\braket{\xi}{\xi}}  (\xi^{\dagger} )^n \xi^n = {\cal C}_0^{n} |\lambda_{n}\lambda_{n-1}...\lambda_1|   \: \: \:  \: \:  \: \:  \: \: \:  \: \:  \: \:  \: \: \:  \: \:  \: \:  \: \: \:  \: n^{th} \: Moment\:  \: \: \nonumber \\
... \nonumber \\
\int d \xi^{\dagger} d \xi \frac{1}{\braket{\xi}{\xi}}  (\xi^{\dagger})^{N-1} \xi^{N-1} = {\cal C}_0^{N-1}| \lambda_{N-1} \lambda_{N-2}...\lambda_1|  \: \: \:  \: \: (N-1)^{th} \: Moment \:  \: \:  
\end{eqnarray}
which is consistent with
\begin{eqnarray}
\int d \xi^{\dagger} d \xi \: \: 1 = 0\nonumber \\
\int d \xi^{\dagger} d \xi  \: \: \xi^{\dagger} \xi = 0 \nonumber \\
\int d \xi^*{\dagger} d \xi  \: \:  (\xi^{\dagger})^2 \xi^2= 0 \nonumber \\
... \nonumber \\
\int d \xi^{\dagger} d \xi \: \:  (\xi^{\dagger})^{N-1}  \xi^{N-1} = {\cal C}_0^{N-1} |\lambda_{N-1} \lambda_{N-2}... \lambda_1|
\end{eqnarray}

While the first (moment) variety of integral can be checked for the limiting fermion and boson cases, the second variety of integrals (without the normalization) cannot be properly defined in the bosonic cases since it isn't convergent. The first method can be treated as a regularized integral.

The one-variable version of these integrals can be defined, in consistency with the above equations, as 
\begin{eqnarray}
\int d \xi \: \xi = 0 \nonumber \\
\int d \xi  \:  \xi^2 = 0 \nonumber \\
.
.
. \nonumber \\
\int d \xi  \: \xi^{N-1} =  {\cal C}_0^{N-1} \sqrt{|\lambda_{N-1} \lambda_{N-2} ...\lambda_1|} \nonumber \\
\int d \xi^{\dagger}  \: (\xi^{\dagger})^{N-1} ={\cal C}_0^{N-1}  \sqrt{|\lambda_{N-1} \lambda_{N-2} ...\lambda_1|} 
\end{eqnarray}

It is possible to define an identity operator, so that (using Equation (29)),
\begin{eqnarray}
\mathcal I = \int d \xi^{\dagger} d \xi  \: \frac{1}{\braket{\xi}{\xi}}  \ket{{\cal C}_0^{-1} \xi^{\dagger}} \bra{\xi^{\dagger}}\: \: \: \: \: \: \: \: \: \: \: \: \: \: \: \: \: \: \: \: \: \: \: \: \: \: \: \: \: \: \: \: \: \: \: \: \: \: \:  \nonumber \\
= \int d \xi^{\dagger} d \xi  \: \frac{1}{\braket{\xi}{\xi}}  \bigg( \ket{\lambda_0}\bra{\lambda_0} + \frac{{\cal C}_0^{-1} \xi^{\dagger} \xi}{|\lambda_1|} \ket{\lambda_1}\bra{\lambda_1}+ \frac{({\cal C}_0^{-1} \xi^{\dagger} {\cal C}_0^{-1} \xi^{\dagger}) \xi^2}{|\lambda_2 \lambda_1|} \ket{\lambda_2}\bra{\lambda_2} + ... \nonumber \\
+  \frac{({\cal C}_0^{-1} \xi^{\dagger}...{\cal C}_0^{-1} \xi^{\dagger}) \xi^{N-1}  }{|\lambda_{N-1} ...\lambda_1|} \ket{\lambda_{N-1}}\bra{\lambda_{N-1}} \bigg) \nonumber \\
\rightarrow \mathcal I= \ket{\lambda_0}\bra{\lambda_0} +\ket{\lambda_1}\bra{\lambda_1} +...+ \ket{\lambda_{N-1}}\bra{\lambda_{N-1}} \: \: \: \: \: \: \: \: \: \: \: \: \: 
\end{eqnarray}

\section{The Trace and Path Integrals}

The trace of an operator $A$ that commutes with and can be taken through $\xi$ is 
\begin{eqnarray}
Tr({\bf A}) = \int d \xi^{\dagger} d \xi \frac{1}{\braket{\xi}} \bra{{\cal C}_0^{-1} \xi} {{\bf A}} \ket{\xi} \: \: \: \:  \: \: \: \:  \: \: \: \:  \: \: \: \:  \: \: \: \:  \: \: \: \:  \: \: \: \:  \: \: \: \:  \: \: \: \:  \: \: \: \:  \: \: \: \:  \: \: \: \:  \: \: \: \:  \nonumber \\
= \int d \xi^{\dagger} d \xi \frac{1}{\braket{\xi}} \: (   \bra{\lambda_{0}}+ \frac{{\cal C}_0^{-1} \: \xi^{\dagger}}{\sqrt{\lambda_1^*}} \bra{\lambda_1} + \frac{({\cal C}_0^{-1} \: \xi^{\dagger})({\cal C}_0^{-1} \: \xi^{\dagger})}{\sqrt{\lambda_2^* \lambda_1^*}} \bra{\lambda_2} +
 ... +  \frac{({\cal C}_0^{-1} \: \xi^{\dagger})...({\cal C}_0^{-1} \: \xi^{\dagger})}{\sqrt{\lambda_{N-2}^* \lambda_{N-3}^*...\lambda_1^*}} \bra{\lambda_{N-2}} + \nonumber \\
 \frac{({\cal C}_0^{-1} \: \xi^{\dagger})...({\cal C}_0^{-1} \: \xi^{\dagger})}{\sqrt{\lambda_{N-1}^*\lambda_{N-2}^*...\lambda_1^*}} \bra{\lambda_{N-1}} ) {\bf A}  (   \ket{\lambda_{0}}+ \frac{\xi}{\sqrt{\lambda_1}} \ket{\lambda_1} +\frac{\xi^2}{\sqrt{\lambda_2 \lambda_1}} \ket{\lambda_2} + \: \: \: \:  \: \: \nonumber \\
... + \frac{\xi^{N-2}}{\sqrt{\lambda_{N-2} \lambda_{N-3}...\lambda_1}} \ket{\lambda_{N-2}} +
 \frac{\xi^{N-1}}{\sqrt{\lambda_{N-1}\lambda_{N-2}...\lambda_1}} \ket{\lambda_{N-1}} ) \nonumber \\
  = \bra{\lambda_0} {{\bf A}} \ket{\lambda_0} +\bra{\lambda_1} {{\bf A}} \ket{\lambda_1} +...+ \bra{\lambda_{N-1}} {{\bf A}} \ket{\lambda_{N-1}} \: \: \: \:  \: \: \: \:  \: \: \: \:  \: \:  \: \: \:  \: \:  \: \:  \: \: \:  \: \: \:  \: \:  \: \:  \:
\end{eqnarray}
where we have used the regularized integrals from Equation (29).

The action integral is, for a ``Hamiltonian'' $\frac{\mathcal H}{k_B T} \equiv \frac{\epsilon}{k_B T} \alpha^{\dagger} \alpha = {\mathcal K} \epsilon \alpha^{\dagger} \alpha$,
\begin{eqnarray}
\mathcal Z = Tr \left( e^{- {\mathcal K} \epsilon \alpha^{\dagger} \alpha} \right)
= \int d \xi^{\dagger} d \xi \frac{1}{\braket{\xi}}  \bra{{\cal C}_0^{-1} \xi} \:  e^{-{\mathcal K} \epsilon \alpha^{\dagger} \alpha} \ket{\xi} 
\end{eqnarray}
The bracketed term can be ``split'' into sub-integrals using the identity operator from Equation (40). We define $\xi_N={\cal C}_0^{-1} \xi_0$ and $\delta \tau = \frac{{\mathcal K}}{N}$.
\begin{eqnarray}
\bra{{\cal C}_0^{-1} \: \xi} \:  e^{- {\mathcal K} \epsilon \alpha^{\dagger} \alpha} \ket{\xi}  = \bra{\xi_N} e^{- \delta \tau \: \epsilon \: \alpha^{\dagger} \alpha} \bigg( \prod_{j=1}^{j=N-1} \int d\xi^{\dagger}_j d\xi_j  \frac{1}{\braket{\xi_j}}\ket{{\cal C}_0^{-1}\: \xi_j^{\dagger}} \bra{\xi_j^{\dagger}} e^{-  \delta \tau \: \epsilon \: \alpha^{\dagger} \alpha}  \bigg)  \ket{\xi_0}  \nonumber \\
=\prod_{j=1}^{j=N} \int d\xi^{\dagger}_j d\xi_j   \exp{- \ln{  \bra{\xi_{j}}\ket{\xi_{j}}  } + \ln{  \bra{\xi_j^{\dagger}}\ket{{\cal C}_0^{-1} \: \xi_{j-1}^{\dagger}}  } } e^{- \delta \tau\epsilon \: \xi_{j}  \: {\cal C}_0^{-1} \:  \xi^{\dagger}_{j-1}}  \:  \: \:  \:
\end{eqnarray}
Using the scalar products as defined in Equation (28) and expanding the logarithm to lowest order,
\begin{eqnarray}
\bra{\xi_{j}}\ket{\xi_{j}}  \approx 1 + \xi_{j}^{\dagger} \xi_{j} \nonumber \\
\rightarrow \ln{\bra{\xi_{j}}\ket{\xi_{j}} } \approx \xi_{j}^{\dagger} \xi_{j} \nonumber \\
\bra{\xi_j^{\dagger}}\ket{{\cal C}_0^{-1} \: \xi_{j-1}^{\dagger}}   \approx 1 +  \: \xi_j \: {\cal C}_0^{-1} \: \xi_{j-1}^{\dagger} \approx 1 +  \xi_{j-1}^{\dagger}  \: \xi_j\nonumber \\
\rightarrow \ln{ \bra{\xi_j^{\dagger}}\ket{{\cal C}_0^{-1} \: \xi_{j-1}^{\dagger}} } \approx   \xi_{j-1}^{\dagger} \xi_j \nonumber \\
\delta \tau\epsilon \: \xi_{j}  \: {\cal C}_0^{-1} \:  \xi^{\dagger}_{j-1} = \delta \tau\epsilon\:  \xi^{\dagger}_{j-1} \xi_{j}   
\end{eqnarray}
the integral reduces to
\begin{eqnarray}
\bra{{\cal C}_0^{-1} \: \xi} \:  e^{-\delta \tau \epsilon \beta \alpha} \ket{\xi} 
=\prod_{j=i}^{j=N} \int d\xi^{\dagger}_j  d\xi_j   \exp{- (\xi_{j}^{\dagger} - \xi_{j-1}^{\dagger}) \xi_j -   \delta \tau \: \epsilon \:  \xi^{\dagger}_{j-1}\xi_{j}}  \nonumber \\
\Rightarrow \int D \xi^{\dagger} D \xi  \exp{-  \int d\tau \:  \xi^{\dagger} (- \partial_{\tau} + \epsilon) \xi  }  
\end{eqnarray}
with boundary conditions for $\xi(\tau), \tau \in (0, {\cal K})$ appropriate to $\xi_N = {\cal C}_0^{-1} \xi_0$.

\subsection{Periodicity of $\xi$}

To define boundary conditions, we use the $Tr({\cal C}_0^{-1})$ as the boundary condition. This is consistent with $\xi_0 \rightarrow \xi_N=\xi_0$ for bosons and $\xi_0 \rightarrow \xi_N = - \xi_0$ for fermions \cite{Murayama1}. Hence, defining $v=i \log Tr({\cal C}_0^{-1})$,
\begin{eqnarray}
\xi_N = \xi_0 \: e^{i (i \log Tr({\cal C}_0^{-1}))} = \xi_0 \: e^{i v} 
\end{eqnarray}

We generalize
\begin{eqnarray}
\rightarrow\xi(\tau) =
 \sum_{q=-\infty}^{q=+\infty} \zeta_q e^{i \frac{(2 q \pi-v)}{\mathcal{K}} \tau}
\end{eqnarray}
This leads to the integral for finite-$N$, using the rules we postulated before in Equations (29) and (30)
\begin{eqnarray}
\mathcal Z  = \prod_{q=- \infty}^{q=\infty} \int d\zeta^{\dagger}_q d \zeta_q \exp{ - \zeta_q^{\dagger} \: \left(- i \frac{(2 q-\frac{v}{\pi}) \pi}{\mathcal{K} } + \epsilon \right)\: \zeta_q} \nonumber \\
\propto  \prod_{q=- \infty}^{q=\infty} \left(- i \frac{(2 q-1)\pi-(\frac{v - \pi}{\pi}) \pi}{\mathcal{K} } + \epsilon \right)^{N-1} = \prod_{q=0}^{q=\infty} \bigg[ 1 + \left(\frac{\epsilon + i \frac{v- \pi}{\mathcal K}}{\frac{(2 q-1) \pi}{\mathcal{K} } } \right)^2 \bigg]^{(N-1)} \nonumber \\
= \prod_{q=0}^{q=\infty} \bigg[ 1 + \left(\frac{{\mathcal K} \epsilon +i (v- \pi)}{(2 q-1) \pi } \right)^2 \bigg]^{(N-1)} \nonumber \\
\propto \cosh\left( \frac{ {\mathcal K} \epsilon +i (v- \pi)}{2} \right)^{N-1} \nonumber \\
\end{eqnarray}

In the above, we use the relations ($a$ is real in the below and the third equation is a general version of the second)
\begin{eqnarray}
\prod_{q=1}^{q=\infty} \left(1 + \frac{x^2}{q^2} \right)= \frac{\sinh \frac{\pi x}{2}}{\frac{\pi x}{2}} \nonumber \\
\prod_{q=1}^{q=\infty} \left(1 + \frac{x^2}{(2q-1)^2} \right)= \cosh \frac{\pi x}{2} \nonumber \\
\prod_{q=1}^{q=\infty} \left(1 + \frac{x^2}{(2q-a)^2} \right)= \frac{\Gamma(1 - \frac{a}{2})}{\Gamma(\frac{-a-ix+2}{2}) \: \Gamma(\frac{-a+ix+2}{2})} \nonumber \\
\cosh(a+i b)+ \cosh(a-ib)=\cosh(a) \cos(b) 
\end{eqnarray}

Hence, with $\theta = \pi, N=2$, i.e., for fermions, we get $\mathcal Z \propto \cosh(\frac{ {\mathcal K} \epsilon }{2})$. Separately,
when $\theta=0, N \rightarrow \infty$, i.e., for bosons, the above formula reduces to the expression for bosons, i.e., $\mathcal Z \propto \frac{1}{\sinh(\frac{ {\mathcal K} \epsilon }{2})}$

\section{Differentiation of generalized Grassmann variables}

We wish to replicate the operator $\theta$-commutator $\bigg( \alpha, \alpha^T \bigg)_{\theta}=1$ with coherent state variables. Realizing that $\xi$'s have to be treated as matrices and comparing the situation with Equation (1) and (10), we impose,
\begin{eqnarray}
 \frac{\partial}{\partial \xi}\bigg( \xi f \bigg) = f + z^{-1} \: \xi \frac{\partial}{\partial \xi} f  \nonumber \\
  \frac{\partial}{\partial \xi^T}\bigg( \xi^T f \bigg) = f + z \: \xi^T \frac{\partial}{\partial \xi^T} f  
\end{eqnarray}
so that there is a $z$ or $z^{-1}$ that accompanies switching the derivative and the variable. This is consistent with the bosonic and fermionic case and permits us to deduce the uncertainty relation for the coherent variable and its conjugate momentum, i.e.,  
\begin{eqnarray}
\bigg(  \frac{\partial}{\partial \xi}, \xi  \bigg)_{-\theta} f(\theta) =\frac{\partial}{\partial \xi} (\xi f ) - z^{-1} \: \xi \frac{\partial}{\partial \xi} f = f \nonumber \\
\bigg(  \frac{\partial}{\partial \xi^T}, \xi^T  \bigg)_{\theta} f(\theta) =\frac{\partial}{\partial \xi^T} (\xi^T f ) - z \: \xi^T \frac{\partial}{\partial \xi^T} f = f
\end{eqnarray}
An immediate consequence is
\begin{eqnarray}
 \frac{\partial}{\partial \xi} \xi^n = \left(  1 + z^{-1} +  z^{-2} + ... + z^{-(n-1)} \right) \xi^{n-1} \nonumber \\
  \frac{\partial}{\partial \xi^T} (\xi^T)^n = \left(  1 + z^{1} +  z^{2} + ... + z^{n-1} \right) (\xi^T)^{n-1}
 \end{eqnarray}
 The above equation is consistent on both sides if we were to set $n=N$, for $\xi^N=(\xi^T)^N = 0$, as well as $1 + z^{1} +  z^{2} + ... + z^{N-1}=0$ and $1 + z^{-1} +  z^{-2} + ... + z^{-(N-1)}=0$.
 
We are going to assume that $\xi$ and $\xi^T, \xi^{\dagger}$ are all independent of each other.
By demanding consistency as in
\begin{eqnarray}
 \frac{\partial}{\partial \xi}  \xi \xi^T = \xi^T =  \frac{\partial}{\partial \xi} z   \xi^T \xi  \nonumber \\
  \frac{\partial}{\partial \xi^T}  \xi^T \xi = \xi =  \frac{\partial}{\partial \xi^T} z^{-1}   \xi  \xi^T
\end{eqnarray}
 we deduce the rules for switching the partial derivative and the transposed variable,
 \begin{eqnarray}
\frac{\partial}{\partial \xi} ( \xi^T f) = z^{-1}\xi^T \frac{\partial}{\partial \xi}f \nonumber \\
\frac{\partial}{\partial \xi^T} ( \xi f) = z\xi \frac{\partial}{\partial \xi^T}f
 \end{eqnarray}
Also,
 \begin{eqnarray}
\frac{\partial}{\partial \xi} e^{-\xi \xi^{\dagger}} = -  \xi^{\dagger} e^{-\xi \xi^{\dagger}} \nonumber \\
\frac{\partial}{\partial \xi^{\dagger}} e^{-\xi \xi^{\dagger}} = -  {\cal C}_0 \: \xi e^{-\xi \xi^{\dagger}}
\end{eqnarray}

In addition, by considering switching the order of partial derivatives, we get
\begin{eqnarray}
\frac{\partial }{\xi^T} \frac{\partial }{\partial \xi} = z^{-1}  \frac{\partial }{\partial \xi} \frac{\partial }{\xi^T}
\end{eqnarray}
which is derived from
\begin{eqnarray}
\frac{\partial }{\xi^T} \frac{\partial }{\partial \xi} \:  \xi \xi^T=1 =  z^{-1}  \frac{\partial }{\partial \xi} \frac{\partial }{\xi^T} \: z \:  \xi^T \xi 
\end{eqnarray}

\section{2-d and 1-d harmonic oscillator: Generalized Hermite polynomials}

Let's solve for the two-dimensional oscillator first. The operators $\alpha, \alpha^T$ are the usual annihilation and creation operators. Let's assume ${\cal A}, \Gamma, \Upsilon$ are all complex numbers that commute with the $\xi, \xi^T$.
\begin{eqnarray}
\alpha = \frac{{\cal A}}{\sqrt{2}} \left( \xi + \Gamma \frac{\partial}{\partial \xi^T} \right)   \nonumber \\
\alpha^T = \frac{{\cal A}}{\sqrt{2}} \left( \xi^T  + \Upsilon  \frac{\partial}{\partial \xi} \right)
\end{eqnarray}
If we want this to be consistent with
\begin{eqnarray}
\alpha \alpha^T - z \alpha^T \alpha = 1
\end{eqnarray}
we derive the simplest solution, that matches the conditions for the case of fermions as well as bosons, i.e.,
\begin{eqnarray}
\Gamma = 1 \nonumber \\
\Upsilon = - \frac{1}{z} \nonumber \\
{\cal A} = 1
\end{eqnarray}
i.e., 
\begin{eqnarray}
\alpha = \frac{1}{\sqrt{2}} \left( \xi + \frac{\partial}{\partial \xi^T} \right)   \nonumber \\
\alpha^T = \frac{1}{\sqrt{2}} \left( \xi^T  - z^{-1}  \frac{\partial}{\partial \xi} \right)
\end{eqnarray}
The ground state wave-function $f_0$ is found from $\alpha f_0 = 0$, which leads to
\begin{eqnarray}
\frac{1}{\sqrt{2}} \left( \xi +  \frac{\partial}{\partial \xi^T} \right) f_0 = 0 \nonumber \\
\rightarrow f_0 \sim  (\xi_T)^m e^{-  \xi^T \xi} =  (\xi_T)^m e^{- z^{-1} \xi \xi^T}
\end{eqnarray}
Note that upon expanding the exponential multiplying by the polynomial in $\xi)T$, we'd keep terms up to $\xi^{N-1}, (\xi^T)^{N-1}$ as higher powers are $0$.
We can construct higher wavefunctions using the creation operator, $f_1 = \alpha^T f_0$ etc.

To obtain the wave-function for the 1-d harmonic oscillator, it is reasonable to assume (in line with the method one uses with the bosonic case\cite{Wess}) that $\alpha, \alpha^T$ are real. In addition, to not allow powers of $\xi^T$ in the ground-state function, we set $\xi = \xi^T$ . We then deduce
\begin{eqnarray}
\alpha = \frac{1}{\sqrt{2}} \left( \xi + \frac{\partial}{\partial \xi} \right)   \nonumber \\
\alpha^T = \frac{1}{\sqrt{2}} \left( \xi  -\cos \theta \:   \frac{\partial}{\partial \xi} \right)
\end{eqnarray}

which yields, when one carries out the above construction, terms that reduce to Hermite polynomials (albeit with series and exponentials terminated at $x^{N-1}, y^{N-1}$), i.e.,
\begin{eqnarray}
f_0 \sim C e^{- \frac{\xi^2}{2}} \nonumber \\
f_1 \sim \frac{1}{\sqrt{2}}C (1+\cos \theta) \xi e^{- \frac{\xi^2}{2}} \nonumber \\
f_2 \sim \frac{1}{\sqrt{2}}C (1+\cos \theta) (\xi^2 (1+\cos \theta) - \cos \theta) e^{- \frac{\xi^2}{2}} 
\end{eqnarray}
We note that this construction does not work for $\theta=\pi$, the $\alpha$'s cannot be real.

\section{Wave-function for two anyons}

The excitations described in this paper represent one of the two ways two anyons can be braided amongst each other. While we have yet to construct a description for anyons here, we could consider start by studying ``overons'' (as opposed to ``underons'' which have the complex conjugate eigenvalues).

For two dimensions, we had, for the ground state for overons (as well as for underons)
\begin{eqnarray}
f_0 = C \:  (\xi^T)^m \: e^{- \xi^T \xi}
\end{eqnarray}
which are the usual holomorphic functions.

For two excitations, the wave-function needs to possess the proper symmetry upon exchange, hence would be
\begin{eqnarray}
f(\xi_1, \xi_1^T, \xi_2, \xi_2^T) = C (\xi_1^T + z \xi_2^T)^N e^{-\xi_1^T \xi_1 - \xi_2^T \xi_2}
\end{eqnarray}
This works as the exchange $1 \leftrightarrow 2$ causes $(\xi_1^T + z \xi_2^T)^N$ to go to
\begin{eqnarray}
(\xi_1^T + z \xi_2^T) \rightarrow (\xi_2^T + z \xi_1^T)^N  = (\xi_1^T + z^{-1} \xi_2^T)
\end{eqnarray}
which produces a wave-function for the excitation with the opposite exchange characteristic, i.e., for ``underons''. While this would not be a possible symmetry for ``overons'', it could represent an appropriate anyon wave-function.

However, as can be quickly checked, the overlap of this function with the Laughlin-like \cite{Laughlin} alternative
\begin{eqnarray}
f_{\cal L}(\xi_1, \xi_1^T, \xi_2, \xi_2^T) = C (\xi_1^T - \xi_2^T)^N e^{-\xi_1^T \xi_1 - \xi_2^T \xi_2}
\end{eqnarray}
is {\bf  non-zero only for odd $N$}, as only terms with both $\xi^T \xi$ raised to powers gives non-zero results.

This is clear, as looking at individual non-zero terms in the overlap integral, i.e.,
\begin{eqnarray}
\int d \xi_1^T d \xi_1  d \xi_2^T d \xi_2 \: (\xi_1^T + z \xi_2^T)^N   \:  (\xi_1^T - \xi_2^T)^N \:  e^{- 2 \xi_1^T \xi_1 - 2 \xi_2^T \xi_2} \nonumber \\
\propto \sum_{m=0}^{N-1} (\xi_1^T)^m (z \xi_2^T)^{N-m} (\xi_1)^m (- \xi_2)^{N-m} \: (- \xi^T_1 \xi_1)^{N-m} (- \xi_2^T \xi_2 )^m \nonumber \\
= \sum_{m=0}^{N-1} (-z)^{-m}  = \frac{1 - (-1)^N}{1 - \frac{1}{z}}
\end{eqnarray}
which shows that the overlap integral is 0 for even-$N$ and proportional to ($= \frac{2}{1 - \frac{1}{z}}$) for odd-$N$. The overlap is of order unity for small odd-$N$, which explains why the function works so well, even for particles with fractional statistics.

\section{Remarks on the propagation of Anyons}

In the usual field theory of fermions \cite{Banks, QFT}, the evolution of a two-particle state is expressed as a perturbative series, starting with the two ``bare'' propagators followed by successively more complex interactions, involving vertices of the interactions and loops between such vertices. 

Let's say we start with two identical fermions ($1$ and $2$), treated as a product of fairly well-separated wave-functions, at points $A$ or $B$. Starting at these separated spots, they can be end up at two widely-separated spots with in all possible ways - either through direct propagation, or with crossed propagation, i.e.,
\begin{eqnarray}
\psi_{Total}(x_1, x_2) = \psi_A(x_1) \psi_B(x_2) - \psi_A(x_2) \psi_B(x_1)
\end{eqnarray}

Usually, the wave-function with exchanged coordinates (the second term) is extremely small, so we can approximate the propagation neglecting the crossed term.

\begin{figure}[h!]
\caption{Two anyon propagation}
\centering
\includegraphics[scale=.5]{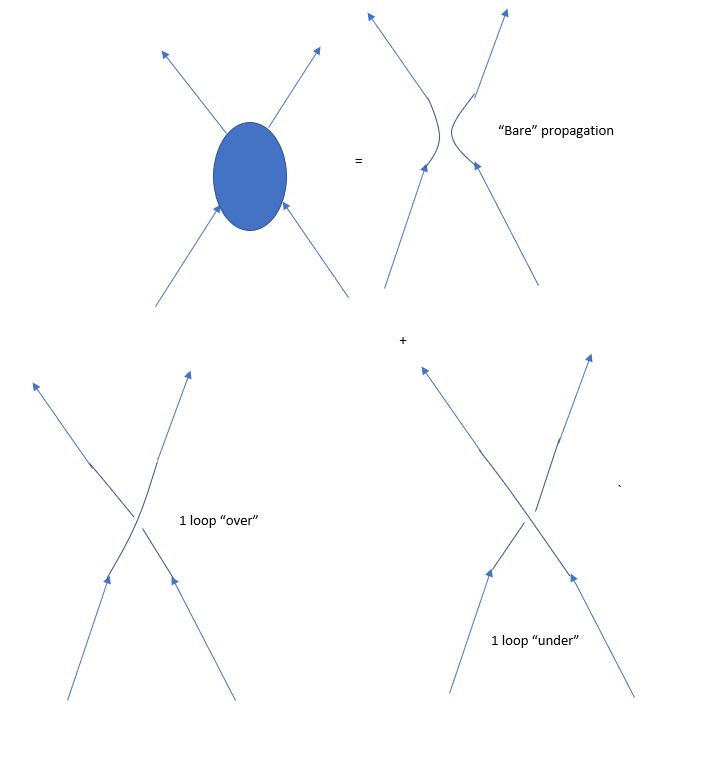}
\includegraphics[scale=.5]{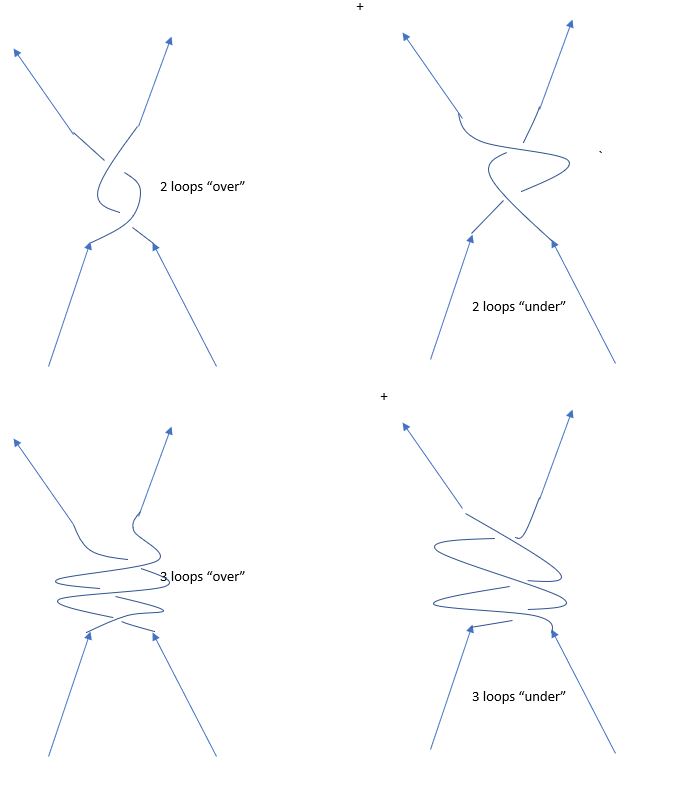}
\end{figure}

The propagator for two anyons, however, has an interesting twist. The propagation of ``bare'' anyons is connected to the value of $\theta$. Two anyons can propagate, as in Fig. 2 while winding around each other 0,2,4,... times (since the final result would then be indistinguishable from no winding) and we need to sum over all these possible alternatives. Suppose $\theta = \frac{2  M \pi}{2 Q+1}$ where $Q$ is an integer and the integers $M,\: 2 Q+1$ are co-prime. Each ``winding'' produces a multiplicative factor of ${\hat \sigma} = e^{i \theta}$ into the amplitude for the two-anyon propagator. Since we need to sum over all the separate ways this can happen, we have to restrict the number of windings to be less than $2Q+1$, when we revert to zero windings. The following factors are useful to define.
\begin{eqnarray}
a_{1} = 1 +{\hat \sigma}^2 + {\hat \sigma}^4 + ... + {\hat \sigma}^{2 Q} \equiv \frac{e^{-i \frac{\theta}{2}}}{2 \cos \frac{\theta}{2}}\nonumber \\
b_{1}=1 + {\hat \sigma}^{-2} +{\hat \sigma}^{-4} + ... + {\hat \sigma}^{-2 Q} \equiv   \frac{e^{i \frac{\theta}{2}}}{2 \cos \frac{\theta}{2}} \nonumber \\
a_{2} =  {\hat \sigma}+ {\hat \sigma}^3 + ... + {\hat \sigma}^{2 Q+1} \equiv   \frac{e^{i \frac{\theta}{2}}}{2 \cos \frac{\theta}{2}} \nonumber \\
b_{2} = {\hat \sigma}^{-1} + {\hat \sigma}^{-3} + ... +{\hat \sigma}^{-(2 Q+1)} \equiv    \frac{e^{-i \frac{\theta}{2}}}{2 \cos \frac{\theta}{2}} \nonumber
\end{eqnarray}
so that the amplitude for the propagation of two anyons is proportional to $S_1+S_2=1$ since
\begin{eqnarray}
S_{1} = a_{1} + b_{1} \equiv  1 \nonumber \\
S_{2}=a_{2} + b_{2} \equiv  1
\end{eqnarray}

\begin{figure}[h!]
\caption{Odd and Even Roots of Unity}
\centering
\includegraphics[scale=.5]{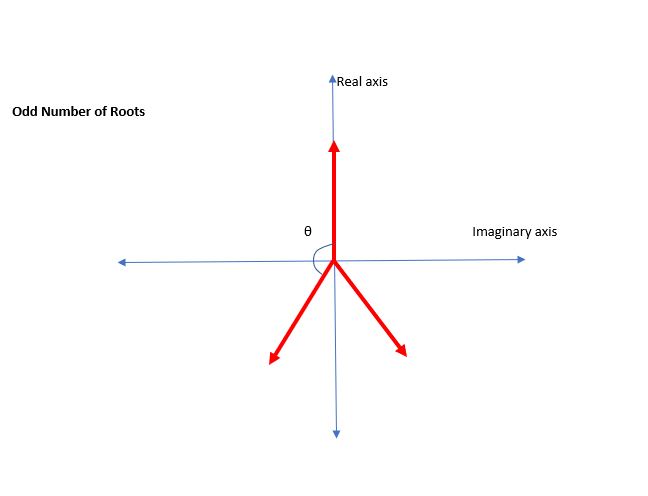}
\includegraphics[scale=.5]{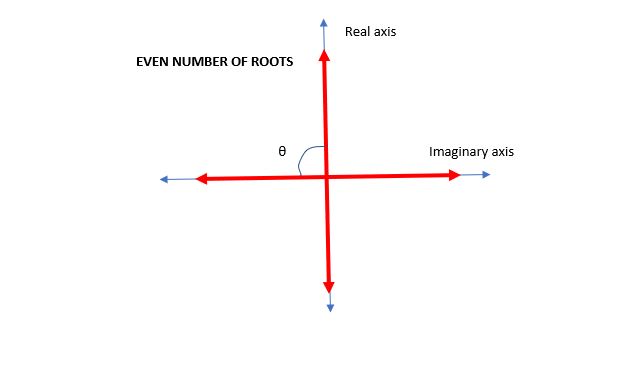}
\end{figure}
On the other hand, if $\theta = \frac{2 M \pi}{2 Q}$, with the caveat $\abs{Q}>1$ and $M, \: 2Q$ co-prime, then the corresponding sums are 
\begin{eqnarray}
p_{1} = 1 +{\hat \sigma}^2 + {\hat \sigma}^4 + ... + {\hat \sigma}^{2 (Q-1)} \equiv 0 \nonumber \\
q_{1}=1 + {\hat \sigma}^{-2} +{\hat \sigma}^{-4} + ... + {\hat \sigma}^{-2 (Q-1)} \equiv  0 \nonumber \\
p_{2} =  {\hat \sigma}+ {\hat \sigma}^3 + ... + {\hat \sigma}^{2 Q-1} \equiv  0 \nonumber \\
q_{2} = {\hat \sigma}^{-1} + {\hat \sigma}^{-3} + ... +{\hat \sigma}^{-(2 Q-1)} \equiv   0 \nonumber
\end{eqnarray}
so that the amplitude for the propagation is $\frac{1}{2}(R_1+R_2) = 0$ below since
\begin{eqnarray}
R_{1} = p_{1} + q_{1} \equiv  0 \nonumber \\
R_{2}=p_{2} + q_{2} \equiv  0
\end{eqnarray}
The exclusion of $Q=1$ (in the `even' case above) is also easy to see (this explicitly means two fermions can indeed propagate). Note that if $Q=1$, the individual terms $p_{1},q_{1},p_{2} ,q_{2}$ are all $1$ or $-1$. There are no cancellations.

Anyons with even-denominator $\theta$ cannot propagate freely, purely from geometrical considerations, while those with odd-denominator $\theta$ can.

This result might seem surprising, but it is clear from studying the roots of unity on the complex plane. Consider Fig. 3, where the odd and even roots of unity are displayed for the case of 3 and 4, respectively. It is clear, for instance, that the sum $1 + e^{2 i \theta}$ is non-zero in the odd-case, while the sum $1+e^{2 i \theta}$ is zero in the even case.

The factors $S_{1}, S_{2}, R_{1}, R_{2}$ multiply the total amplitudes, assuming there are no energetic consequences within the action to meandering paths that wind around each other multiple number of times. It would, therefore, not be surprising that multiple anyon propagation is suppressed with even-denominator theta. It is, however, possible that higher order interaction terms will allow propagation for even-denominator-theta anyons. This will have physical consequences for transport and localization of states with odd and even denominator $\theta$.

This behavior (for odd and even-denominator $\theta$) is quite robust to the addition of a small cost for extra windings. Suppose the anyons encountered a cost, a factor of $h = 1 - \delta$, small $\delta$ cost,  for each extra winding around the other anyon. Then the corresponding sums would become, for {\underline{even}}-denominator $\theta= \frac{2 M \pi}{2 Q}$,
\begin{eqnarray}
\hat p_{1} = 1 +  h^2 {\hat \sigma}^2 + h^4 {\hat \sigma}^4 + ... + (h^2)^{Q-1}{\hat \sigma}^{2 (Q-1)} \approx \delta \frac{2Q}{1 - e^{2 i \theta}} \nonumber \\
\hat q_{1}=1 + h^2 {\hat \sigma}^{-2} +h^4 {\hat \sigma}^{-4} + ... + (h^2)^{Q-1} {\hat \sigma}^{-2 (Q-1)} \approx \delta \frac{2Q}{1 - e^{-2 i \theta}}\nonumber \\
\hat p_{2} =  h{\hat \sigma}+ h^3 {\hat \sigma}^3 + ... + h^{2Q-1} {\hat \sigma}^{2 Q-1}  \approx h e^{i \theta} \delta \frac{2Q}{1 - e^{2 i \theta}} \nonumber \\
\hat q_{2} = h {\hat \sigma}^{-1} + h^3 {\hat \sigma}^{-3} + ... +h^{2Q-1} {\hat \sigma}^{-(2 Q-1)} \approx h e^{-i \theta} \delta \frac{2Q}{1 - e^{-2 i \theta}} \nonumber \\
\hat R_{1} = \hat p_{1} + \hat q_{1} = 2 Q \delta \nonumber \\
\hat R_{2}=\hat p_{2} + \hat q_{2} =0 \: (\: to \: order \: \delta)
\end{eqnarray}
while the odd-denominator $\theta=\frac{2 \pi}{2 Q+1}$ results are of $\sim 1$ to the same order. The suppression is, therefore, robust for small additional cost for anyons ``winding'' around each other.

\section{Conclusions}

We have studied the arithmetic and calculus of overons, excitations under generalized statistics interpolating between fermions and bosons. We have described the fuzzy ``pancake'' surface that best describes the eigenvalue surface for the algebra. Further, the calculus of coherent state variables is studied, as is the partition function for these states. We then proceed to study generalizations of the Hermite polynomials.

Using the results, we have studied some consequences for the field theory of anyons that immediately result from the calculus of coherent state variables as well as from the geometrical interpretation. These demonstrate the appropriateness of the Laughlin wave-function to describe 2-anyon states. In addition, there are rather simple geometrical reasons why even-denominator $\theta$ anyons cannot propagate freely and can only do so in the presence of anyon-anyon interactions.

\section{Acknowledgments}

SR acknowledges the hospitality and intellectual stimulation of the Rutgers Department of Physics \& Astronomy and the NHETC at Rutgers. Much of this work benefited from very useful advice and suggestions from Professor Scott Thomas. He also acknowledges the collaborative atmosphere provided at the ITP, Santa Barbara.

\section{APPENDIX 1: A dynamical system analog for the eigenvalue spectrum}

\subsection{Boson starting point}

Consider the problem of a boson, represented as a scalar field, defined on a circle $(S^1)$ around another boson. The circle is discretized into $N$ points, labelled $0, 1, ..., N-1$. The lattice constant (between the discrete points) is $a = \frac{2 \pi}{N} = \theta$.
The potential energy part of the Hamiltonian, after partial integration, leads to 
\begin{eqnarray}
{\cal H}_P^{boson} = -  \sum_{j=0}^{N-1} {\cal K} \frac{ \phi(j)}{2} \left( \frac{\phi(j+1)-\phi(j)}{a} - \frac{\phi(j) - \phi(j-1)}{a} \right)
\end{eqnarray}
Transform to Fourier coordinates
\begin{eqnarray}
\phi_j = \frac{1}{\sqrt{N}} \sum_{k = 0}^{N-1} {\tilde \phi}(k) \: e^{i \frac{2 \pi}{Na} k j a} \nonumber \\
\rightarrow {\cal H}_P^{boson} = \frac{{\cal K}}{a} \sum_{k=0}^{N-1} {\tilde \phi}(k) {\tilde \phi}(-k) \: (1 - \cos (\frac{2 \pi}{Na} k a)) \nonumber \\
= 2 \frac{{\cal K}}{a} \sum_{k=0}^{N-1} {\tilde \phi}^*(k) {\tilde \phi}(k) \:\sin^2(\frac{k \theta}{2})
\end{eqnarray}

\subsection{Fermion starting point}

Start with a fermion, represented by a complex field defined on the same circle. The Hamiltonian would have a first-order derivative, as below
\begin{eqnarray}
{\cal H}_P^{fermion} = -i  \sum_{j=0}^{N-1} \frac{{ \cal K}}{2} \phi^*(j) \left( \frac{\phi(j)- \phi(j-1)}{a} \right) + c.c.
\end{eqnarray}
Again, writing this in Fourier space, using $a = \frac{2 \pi}{N}$,
\begin{eqnarray}
{\cal H}_P^{fermion} =  \frac{{\cal K}}{a} \sum_{k=0}^{N-1} {\tilde \phi}^*(k) {\tilde \phi}(k) e^{\frac{-ika}{2}} \sin( \frac{ka}{2}) +c.c. =  \frac{2{\cal K}}{a} \sum_{k=0}^{N-1} {\tilde \phi}^*(k) {\tilde \phi}(k) \cos(\frac{k \frac{2\pi}{N}}{2}) \sin( \frac{k \frac{2\pi}{N}}{2})\nonumber \\
= \frac{{\cal K}}{a} \sum_{k=0}^{N-1} {\tilde \phi}^*(k) {\tilde \phi}(k) \sin(k \theta) \: \: \: \: \: \: \: \: \: \: \: \: 
\end{eqnarray}

Fermion doubling - this would go away if we set $a = \frac{\pi}{N}$.

\subsection{A composite particle starting point}

Composing the above Hamiltonians, choosing to appropriately interpolate between the boson $(\theta=0)$ and fermion $(\theta=\pi)$ cases, we write the discretized version of the fractional ($\nu$) derivative as follows (assume wrap-around coordinatization for a circle)
\begin{eqnarray}
{\cal H}_P^{composite} =  (-i)^{\nu} \frac{{\cal K}}{a^{\nu}} \sum_{j=0}^{N-1} \phi^*(j) \bigg( \phi(j) +(-1) \frac{\Gamma(\nu+1)}{1! \Gamma(\nu)} \phi(j-1) +(-1)^2 \frac{\Gamma(\nu+1)}{2! \Gamma(\nu-1)} \phi(j-2) \nonumber \\
+...+(-1)^{N-1} \frac{\Gamma(\nu+1)}{(N-1)!\Gamma(\nu-N+1)} \phi(j-(N-1)) \bigg) + c.c. \: \: \: \: \: \: \: \: \: \: \: \: \: \: \: \: \: \: \: \: \: \: \: \: 
\end{eqnarray}
which, in Fourier space becomes
\begin{eqnarray}
{\cal H}_P^{composite} =  (-i)^{\nu} \frac{{\cal K}}{a^{\nu}} \sum_{k=0}^{N-1} {\tilde \phi^*(k)} {\tilde \phi(k)} \bigg( 1 +(-1) \frac{\Gamma(\nu+1)}{1! \Gamma(\nu)} e^{-ika} +(-1)^2 \frac{\Gamma(\nu+1)}{2! \Gamma(\nu-1)} e^{-2ika} \nonumber \\
+...+(-1)^{N-1} \frac{\Gamma(\nu+1)}{(N-1)!\Gamma(\nu-N+1)} e^{-(N-1)ika} \bigg) + c.c. \: \: \: \: \: \:  \nonumber \\
= (-i)^{\nu} \frac{{\cal K}}{a^{\nu}} \sum_{k=0}^{N-1} {\tilde \phi^*(k)} {\tilde \phi(k)} \bigg( 1 - e^{-ika}\bigg)^{\nu}  \: \: \: \: \: \:   \nonumber \\
=4^{\nu} \frac{{\cal K}}{a^{\nu}} \cos \frac{ka \nu}{2} \bigg( \sin\frac{ka}{2} \bigg)^{\nu} \: \: \: \: \: \:  \: \: \: \: \: \:  \: \: \: \: \: \:  \: \: \: \: \: \:  \: \: \: \: \: \:  \: \: \: \: \: \:  
\end{eqnarray}

This is a version of fermion doubling for the composite particles.

\end{document}